# Strongly coupled single quantum dot in a photonic crystal waveguide cavity

F.S.F. Brossard[1*], X. L. Xu[1*], D.A. Williams[1], M. Hadjipanayi[2], M. Hopkinson[3], X. Wang[2] and R.A. Taylor[2]

[1]Hitachi Cambridge Laboratory, Hitachi Europe Ltd., J. J. Thomson Avenue, Cambridge, CB3 0HE, United Kingdom

[2]Department of Physics, University of Oxford, Parks Road, Oxford, OX1 3PU, United Kingdom

[3]Department of Electronic and Electrical Engineering, University of Sheffield, Mapping Street. Sheffield, S1 3JD, United Kingdom

[*] These authors contributed equally to this work.

## Abstract

Cavities embedded in photonic crystal waveguides offer a promising route towards large scale integration of coupled resonators for quantum electrodynamics applications. In this letter, we demonstrate a strongly coupled system formed by a single quantum dot and such a photonic crystal cavity. The resonance originating from the cavity is clearly identified from the photoluminescence mapping of the out-of-plane scattered signal along the photonic crystal waveguide. The quantum dot exciton is tuned towards the cavity mode by temperature control. A vacuum Rabi splitting of ~ 140 µeV is observed at resonance.

Photonic crystal waveguide (PhCWG) cavities provide the highest quality factors ($Q$s) reported to date whilst maintaining a mode volume $V$ of the order of $(\lambda_0/n)^3$ [1,2]. The resulting very large $Q/V$ makes them particularly suitable candidates for studying strong light-matter interactions. In addition, the PhCWG section can be tailored to form large-scale arrays of low-loss coupled resonators with the possibility of in-line input-output coupling as recently demonstrated [3]. The strong coupling regime with a single quantum dot (QD) has been demonstrated in H1 or L3 type cavities which are isolated from the output-interface of the photonic crystal (PhC) by the photonic lattice [4,5]. The demonstration of the strong coupling regime in a cavity embedded in a PhCWG would constitute an important step towards the realization of quantum computing devices based on an array of coupled photonic microcavities [6].

The PhCWG cavity studied here is based on the local modulation of the PhCWG width [2]. It was designed to emit resonantly at the low energy tail of a high dot density sample such as to limit absorption effects due to the QD ensemble and light losses related with fabrication imperfections [7]. The holes were shifted slightly and gradually away from the PhCWG as indicated in figure 1(a) by a distance adapted from [2]. The gradual shift of the holes by a few nm creates a low-loss resonant mode confined by the mode gap of the PhCWG, based on the same principle as that used in a double heterostructure [1]. The device was fabricated by first defining the PhC pattern lithographically in PMMA with a 100 kV VB6 Leica e-beam machine. The pattern was then transferred onto a 180 nm thick GaAs slab containing InAs QDs using reactive ion etching (RIE) with a $SiCl_4$/Ar mixture, the remaining resist was removed with a $NH_4OH$/acetone soak before HF treatment forming an air-bridge. Finally, a digital etch step was employed using $O_2$/HCl to remove any damaged native oxide resulting from the RIE procedure [8]. An SEM of the fabricated cavity is shown in figure 1(a) and a view of the entire PhCWG structure with no apparent disorder shown in figure 1(b). The expected cavity mode profile, $Q$ and mode volume were obtained using a freely available 3D finite difference time domain (FDTD) package MEEP [9]. A $Q$ of about $2\times10^7$ and a $V\sim1.3\ (\lambda_0/n)^3$ for $n = 3.46$ [10] was found using flux planes surrounding the PhCWG with the same size as the fabricated device. The electric field energy distribution shown at the centre of the cavity in figure 1(c) confirms that such a design should support an ultra-high $Q$ cavity mode.

PL measurements were carried out on a high InAs dot density ($\sim$100 dots/$\mu m^2$) wafer with dots grown by molecular beam epitaxy giving a broad emission centered at 1.3 eV. The sample was mounted in a He-flow cryostat cooled to 5 K and the dots pumped with a $\sim$ 1 μm spot size He-Ne laser obtained using a 100× microscope objective (N.A. 0.75). The emission light was collected by the same objective, dispersed through a 0.55 m spectrometer and detected with a cooled charge-coupled device camera. A typical PL spectra obtained with this cavity is shown in figure 2(a) under a relatively high pumping power of 32 μW. The spectra consisted of an isolated sharp peak and a series of closely spaced peaks at slightly higher energies. These features were reproduced in all the PhCWGs investigated, with peak energies determined by the PhC lattice constant, hence indicating resonances associated with the band structure of the device. The cavity resonance shown by the arrow in figure 2(a) was clearly identified by shifting the laser spot laterally along the PhCWG and through the cavity. We attribute the sharp peaks observed either side of the cavity to Fabry Perot resonances caused by reflections from the cavity-PhCWG interface and PhCWG ends (section A and C of figure 1b) [11].

Dozens of cavities were investigated with various hole sizes $r$, lattice constant $a$ and ratio $r/a$. A snapshot of our results is reported in figure 2(b) showing clear trend towards lower $Q$ for devices emitting at higher energies as a result of cavity linewidth broadening. The $Q$s reported here are orders of magnitude lower than the ultra high values shown elsewhere in Si and GaAs samples for this type of cavity but without QDs [2,12]. We believe some of the losses are associated with absorption by the QD ensemble whose main emission peak is at 1.3 eV and are mostly unavoidable in a high dot density sample [13]. Additional losses could also be caused by fabrication imperfections resulting from the dry etching process which would become particularly severe at higher energies due to polarization mode mixing, as we have recently demonstrated in high $Q$ cavities [7]. Significantly higher $Q$s are expected in these cavities in the future with carefully designed devices and lower QD density samples.

The strong coupling regime was observed with a reduced pump power of 3.2 µW and a QD slightly to the blue of the cavity mode. The QD is strongly red shifted towards the cavity as the temperature increases due to bandgap shrinkage as observed in figure 3(a). In an uncoupled system, both the QD and cavity photon would merge into a single peak without perturbation in their respective quantum states. In contrast, a strongly coupled system is characterized by mixed photon-atom states where the degeneracy in the energy levels of each individual entity is lifted, resulting in the Rabi splitting at resonance. This manifests itself experimentally as two distinct Lorentzian peaks and an anticrossing behaviour when the dot is tuned towards the cavity, as clearly observed in our results shown in figure 3(a). We note that the cavity peak should red shift slightly as the temperature increases due to the increased bulk refractive index of GaAs. The blue shift observed here is attributed to the desorption of condensated gas on the cold sample surface [14]. A Rabi splitting energy $\Delta E$ of about 140 µeV is deduced from the peak positions of the dot and PhCWG cavity mode at resonance (zero detuning) as shown in figure 3(b). Rabi splitting occurs at resonance when the exciton-photon coupling strength becomes greater than the mean of their decay rates so that the energy can be exchanged reversibly between the QD and the cavity mode as expressed by [15]:

$$\Delta E = 2\hbar\sqrt{g^2 - \left(\frac{\gamma_c - \gamma_x}{4}\right)^2} \qquad (1)$$

where $g$ is the exciton-photon coupling constant, $\gamma_c$ and $\gamma_x$ are the cavity and exciton linewidths, respectively. From the full-width at half-maximum of our cavity peak (inset of figure 2a) we measure $\hbar\gamma_c$ = 160 µeV and $\hbar\gamma_x$ = 78 µeV for the single QD with some uncertainty attached to $\hbar\gamma_x$ due to the limited resolution of our spectrometer. We usually have $\hbar\gamma_c \gg \hbar\gamma_x$, besides for the vacuum Rabi splitting we have $\Delta E = 2g$, hence the condition on the observation of strong coupling, taking into account the losses, can be simplified from Eq. (1) to $\Delta E > \hbar \gamma_c /2$ which is clearly the case in our results. The predicted Rabi splitting can be obtained from Eq. (1) by injecting the mode volume $V$ given by the FDTD calculation and an oscillator strength $f = 10.7$ for InAs QDs [15] into the expression for the coupling strength [15]:

$$g = \sqrt{\frac{1}{4\pi\varepsilon_0\varepsilon_r} \frac{\pi e^2 f}{mV}} \qquad (2)$$

where $m$ is the free electron mass, $e$ the electron charge and $\varepsilon_0\varepsilon_r$ the dielectric constant. The results show that the splitting is expected to appear in the PhCWG cavity for relatively modest $Q > 3000$ and to saturate already for $Q \sim 10^4$, as similarly found for the L3 type cavity [16]. The Rabi splitting measured in this work is about 70 % of the 200 μeV predicted for $Q = 8000$. It is clear from previous reported results that the main issue to achieving the maximum splitting is the misalignment between the QD and the antinode(s) of the cavity mode, which causes the coupling strength to decrease [17]. PhC cavities with smaller $V$ than the PhCWG cavity have the potential to achieve larger Rabi splitting but at the expense of more stringent alignment requirements. The present system would benefit greatly from deterministic procedures to locate the position and energy of the quantum dot for the study of coupled high $Q$ PhC resonators for QED applications [18].

In conclusion, we have demonstrated the observation of the strong coupling regime between a single QD and a PhCWG cavity formed by locally tuning the PhCWG lattice. The Rabi splitting reported compares well with previously reported results for L3 and H1 type PhC cavities with smaller mode volume. Future work will investigate whether the state of the exciton-photon coupled system can be probed through the resonances observed in the PhCWG. We believe the results presented here to be an important step towards multiple coupled cavities which can be addressed through the PhCWG for future quantum networks.

*The FDTD computation was performed using the Darwin Supercomputer of the University of Cambridge High Performance Computing Service (http://www.hpc.cam.ac.uk/), provided by Dell Inc. using Strategic Research Infrastructure Funding from the Higher Education Funding Council for England. This research is part of the QIP IRC supported by EPSRC (GR/S82176/01)*

**Figure captions**

Figure 1. (a) SEM of the cavity embedded in a PhCWG with width $W = 0.98\sqrt{3}a$ indicated by the white double arrow, where $a$ is the lattice constant. The holes are shifted within the dashed hexagon by (red arrows) 6 nm, (yellow arrows) 4nm and (blue arrows) 2 nm in a 250 nm PhC lattice with hole size of about 140 nm. (b) zoomed out SEM image of (a) showing the symmetrically defined left (A) and right (C) PhCWG sections and the cavity defined at B. (c) Calculated electric field energy distribution of the cavity mode.

Figure 2. (a) PL spectra of the resonances as the laser spot is shifted through the cavity with a 50 nm resolution XYZ stage attached to the objective. A, B and C correspond to the sections of the PhCWG indicated in figure 1(b). The cavity resonance indicated by the arrow clearly disappears either sides of the cavity. (b) Recorded "$Q$ map" of the cavity resonance for various $a$ and $r/a$.

Figure 3. (a) PL spectra of the strongly coupled QD-cavity mode system for various temperatures in steps of 0.5 K showing the anticrossing as the dot is tuned towards the cavity mode. The arrow in the inset shows the cavity mode with a $Q$ of about 8000, measured at 5K from a Lorentzian fitting shown in red. Two distinct peaks of similar linewidths at resonance can be seen (red spectra). (b) Peak positions of the strongly coupled and uncoupled system for various detunings, showing a Rabi splitting of about 140 μeV at zero detuning. The squares indicate the measured peaks from the strongly coupled system in (a) whilst the red and green lines are obtained from an uncoupled dot and cavity measured on the same sample, respectively. The black lines show the calculated peak positions for a strongly coupled system for a particular detuning energy.

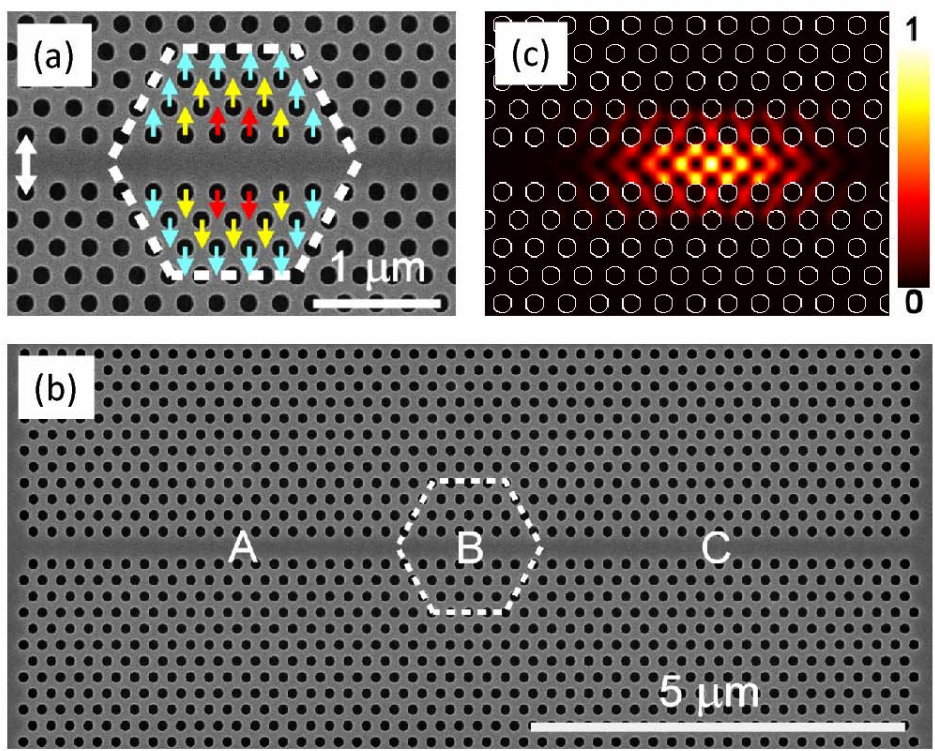

Figure 1

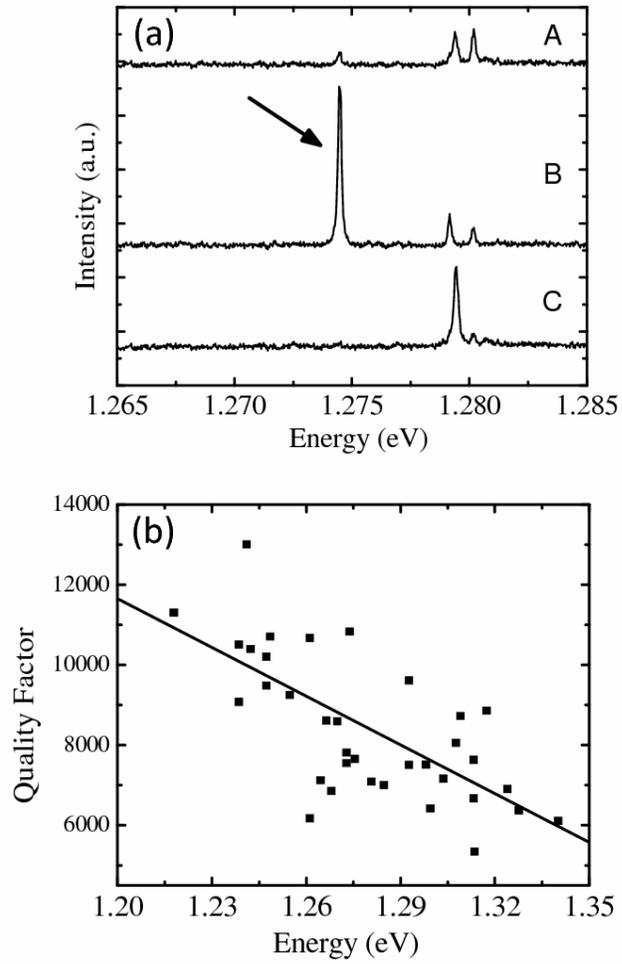

Figure 2

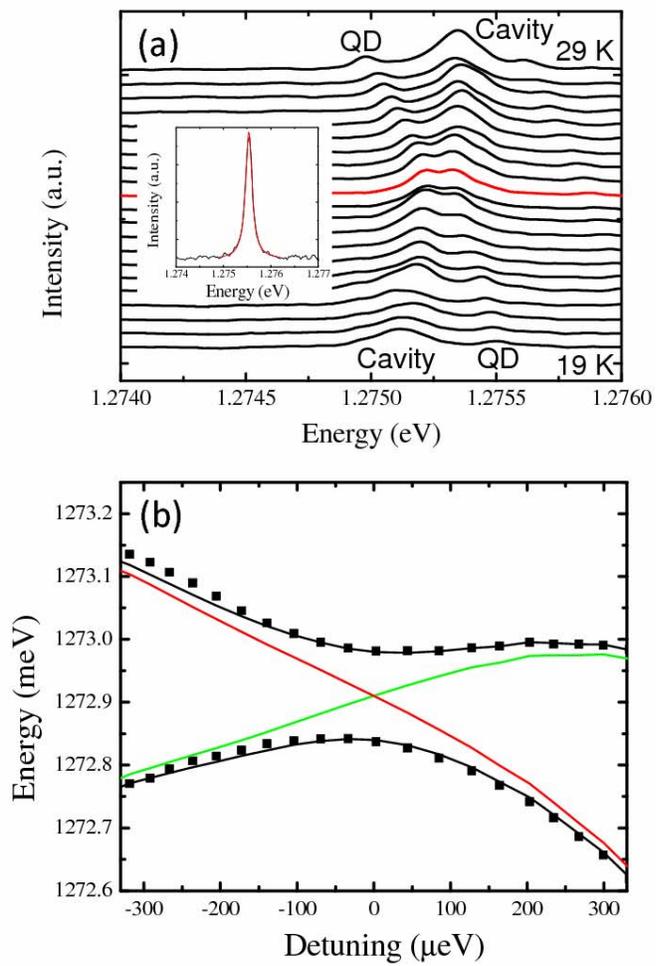

Figure 3